\documentstyle[fleqn,12pt]{article}
\begin{document}
\baselineskip=0.75cm

\begin{titlepage}
\begin{center}
{\Large \bf  An Alternative Approach to Jaynes-Cummings \\Model
with Dissipation at Finite Temperature }
\end{center}

\begin{center}
{\bf    Le-Man Kuang}\\
{\small {\it   Theoretical Physics Dvision, Nankai Institute of
Mathematics,}}\\
{\small {\it   Tianjin 300071, People's Republic of China and}}\\
{\small {\it   Department of Physics and Institute of Physics, Hunan Normal
University,}}\\
{\small {\it   Hunan 410006, People's Republic of China}}\\
{\bf    Guang-Hong Chen and Mo-Lin Ge } \\
{\small {\it   Theoretical Physics Dvision, Nankai Institute of
Mathematics,}}\\
{\small {\it   Tianjin 300071, People's Republic of China }}\\
\end{center}

\begin{abstract}
\baselineskip=0.90cm
An alternative approach to the Jaynes-Cummings model (JCM) with dissipation at
a finite enviromental temperature is presented in terms of a new master
equation under Born-Markovian approximations. An analytic solution of
the dissipation JCM is obtained. A variety of physical quantities  of interest
are
calculated analytically. Dynamical properties of the atom and the field
are investigated in some detail. It is shown that both cavity damping and
environmental temperature strongly affect nonclassical effects in the
JCM, such as collapse and revivals of the atomic inversion, oscillations of
the photon-number distribution, quadrature squeezing of the field and
sub-Poissonian photon statistics.

\end{abstract}
\hspace{0.5cm}
\end{titlepage}
\newpage
\renewcommand{\theequation}{\thesection.\arabic{equation}}
\section{Introduction}
In the past few years there has been considerable interest in
studying the Jaynes-Cummings (JCM) [1] with dissipation since
        the experimental verification of the collapse and revivals of
        Rabi oscillations [2-4]. A number of authors have treated the JCM
        with dissipation by means of analytic approximation [5,6] as
        well as numerical calculations [7-11]. Most of these work dealt
        with the JCM with dissipation only at the zero enviromental
        temperature due to some technical difficulties. As is well
        known, in the JCM one must solve a master equation with cavity
        damping. Generally speaking, it is more difficult to solve
        the master eqution analytically. To our knowledge, no
        analytic solution for this model has been given for general
        initial conditions and environmental temperatures. At zero
        temperature, Agaral and Puri [12] presented an analytic
        solution for the initial state of the field being a vacuum
        state; And Daeubler {\it et al.}. [13] found an analytic expression
        for the atomic inversion and the intensity of the cavity field
        for the initial state of the field being a coherent state by
        using the quasiprobability-distribution technique [14]. At a finite
        temperature, only some numerical solutions [8,9,11] have been
        reported. The purpose of the this paper is intend to present an
        alternative approach to the dissipation JCM at a finite
        temperature by introducing a new master equation which comes
        from a simple coupling between the system and its environment. We
        study  dynamical properties of the field in the
        JCM analytically. In particular, we show that both cavity damping and
        environmental temperature  strongly affect nonclassical effects
        in the JCM, such as collapes and revivals of the atomic
        inversion, oscillations of the photon-number distribution,
        quadrature squeezing of the field, and sub-Poissonian photon
statistics.

              This paper is organized as follows: In Sec.2, a new master
        equation is presented and an explicit expression of the master
        equation for the JCM is obtained. In Sec.3, dynamical properties
        of the atom is investigated. The atomic inversion, the dipole
        moment of the atom and the atomic entropy are calculated
        analytically. Sec.4 is devoted to dynamics of the
        field. Oscillations of the photon-number distribution, quadrature
        squeezing of the field and sub-Poissonian photon statistics are
        discussed in detail. Finally, in Sec.5 we will summarize our
        results and give some concluding remarks.

\section{Master Equation and its Analytic Solution \\ for the JCM}
Let $\hat{H}$ and $\hat{\rho}$ to be the Hamiltonian and the density operator
of
the JCM, respectively. We use a bath of harmonic oscillators to model
the cavity damping which describes the irreversible motion of
$\hat{\rho }$ caused by the enviroment consisting of an infinite set of
 harmonic oscillators. We assume that the atom-field system of the JCM
 the system interacting with the environmemt (or a reservoir)
can be described by the total Hamiltonian
\begin{equation}
\hat{H}_T=\hat{H} + \sum_{i} \hbar \omega_i \hat{b}^+_i\hat{b}_i
+ \hbar \hat{H}\sum_{i} C(\omega_i)(\hat{b}_i+\hat{b}^+_i) +\hbar^2
\hat{H}^2\sum_{i}\frac{|C(\omega _i)|^2}{\omega_i}
\end{equation}
where  $\hat{b_i}$ and $\hat{b}^+_i$ are the boson annihilation and
ceration operators for the enviromemt, in Eq.(2.1)
the second term  is the Hamiltonian of the reservoir,  the third one
 represents  the interaction  between the system and the reservoir
, and the last one is the renormalization term
which compensates for the coupling-induced renormalization of the
potential [15,16]. In the total Hamiltonian  we have adopted a simple
coupling between the system and the reservoir such
that it satisfies the condition \begin{math}[V, H]=0\end{math} ($V$ denotes the
interaction
term in (2.1)), which is required in the back-action
evading and quantum-nondemolition schemes [17] and applied to decoherence
 of quantum system [18-20].

Following the standard master-equation approach [21], making use of the
Hamiltonian (2.1) we can derive the following master equation for the
reduced density operator in the Schr\"{o}dinger picture under
Born-Markovian approximations:
\begin{equation}
\frac{d\hat{\rho} _s(t)}{dt}=\frac{1}{i\hbar}[\hat{H}, \hat{\rho}(t)] - \gamma
[\hat{H}, [\hat{H}, \hat{\rho}(t)]]
- \Delta \omega [\hat{H}, \hat{\rho}(t)]
\end{equation}
where
\begin{equation}
\gamma =\Delta \omega' +  \frac{kT}{\hbar}\lim_{\omega \rightarrow 0}
\frac{J(\omega)|C(\omega)|^2}
{\omega},
\end{equation}
\begin{equation}
 \Delta \omega= \Delta \omega' + 2i {\cal P} \int_{0}^{\infty}{\rm d} \omega
\frac{J(\omega)|C(\omega)|^2}{\omega}
\end{equation}
with
\begin{equation}
\Delta \omega'=i\hbar \int_{0}^{\infty}{\rm d}\omega
\frac{J(\omega)|C(\omega)|^2}{\omega}
\end{equation}
In the above equations, $J(\omega)$ is the spectral density of the
reservoir,  ${\cal P}$ is the Cauchy principal part of the integration
[21], $T$ is the temperature of the enviroment and $k$ is the Boltzmann
constant. In the derivation of the master equation, we have assumed that
the temperature $T$ is high enough so that the Markovian approximation
is valid.

 If we neglect the the Lamb shift term, i.e., the last term on rhs of
 Eq.(2.1), the master equation becomes
\begin{equation}
\frac{d\hat{\rho} (t)}{dt}=\frac{1}{i\hbar}[\hat{H}, \hat{\rho} (t)] -
\gamma[\hat{H}, [\hat{H}, \hat{\rho}(t)]]
\end{equation}
Notoce that this equation  has the same form as the Milburn's equation [22]
under diffusion
approximatiom. However, they come from completely different physical mechanism.
The
former originates from the dissipation while the latter is from the
uncontinuous
and stochastic
unitary evolution. In what follows we will use the master equation (2.6)
to study the JCM with dissipation.

The resonant Jaynes-Cummings Hammiltonian  describing an interaction
of a two-level atom with a single-mode cavity field
 in the rotating-wave approximation is given by
\begin{equation}
\hat{H} = \hbar\omega (\hat{a}^+ \hat{a} + \frac{m}{2} \hat{\sigma}_3)
+\hbar \lambda (\hat{\sigma}_- \hat{a}^+ +\hat{\sigma}_+ \hat{a}),
\end{equation}
where $\omega$ is the resonant  frequency of the cavity field  and the
atomic trasition,  $\lambda$ is the atom-field coupling constant;
$\hat{a}$ and $\hat{a}^+$ are the field annihilation and creation operators,
respectively; $\hat{\sigma}_3$ is the atomic-inversion operator and
$\hat{\sigma}_{\pm}$ are the atomic ``spin flip" operators which satisfy
the relations $[\hat{\sigma}_+ ,\hat{\sigma}_- ]=2\hat{\sigma}_3$ and
$[\hat{\sigma}_3 ,\hat{\sigma}_{\pm}]=\pm 2\hat{\sigma}_{\pm}$. For
simplicity, we take $\hbar =1$ throughout the paper.

We now  find the analytic  solution of the master equation (2.6) applied
to  the Hamiltonian (2.7). Moya-Cessa {\it et al.} [23] have obtained an formal
solution of (2.6) for a Hamiltonian with a little diffrence from (2.7).
In what follows we will present an explicit solution of the master
equation for the Hamiltonian under our consideration.  Following the
approach in refs.[23,24],
 we introduce three  superoperators $\hat{R}$, $\hat{S}$
and $\hat{T}$ which are defined through their actions on the density operator,
respectively,
\begin{eqnarray}
\exp(\hat{R} t) \hat{\rho} (0)&=&\sum^{\infty}_{k=0} \frac{(2\gamma t)^k}
{k!} \hat{H}^k \hat{\rho} (0) \hat{H}^k    \\
\exp(\hat{S} t) \hat{\rho} (0)&=& \exp(-i\hat{H} t) \hat{\rho} (0)
\exp(i\hat{H} t)    \\
\exp(\hat{T} t) \hat{\rho} (0)&=&\exp(- \gamma t \hat{H}^2 )
\hat{\rho} (0) \exp(- \gamma t \hat{H}^2 )
\end{eqnarray}
where  $\hat{\rho} (0)$ is the initial density operator of the atom-field
system, and the Hamiltonia $\hat{H}$ is given by Eq.(2.7).

It can be checked that  the master equation (2.6) has the following
formal solution:
\begin{equation}
\hat{\rho} (t)=\exp(\hat{R} t) \exp(\hat{S} t) \exp(\hat{T} t) \hat{\rho} (0)
\end{equation}

 We assume that initially the field is prepared in the coherent state
$\mid z\rangle $ defined by
\begin{equation}
\mid z\rangle =\sum^{\infty}_{n=0} Q_n \mid n\rangle,
\hspace{1.5cm}
Q_n = exp(-\frac{1}{2} |z|^2 ) \frac{z^n}{\sqrt n!}
\end{equation}
and the atom was prepared in its excited state $\mid e\rangle $, so that the
initial
density operator of the atom-field system takes this form:
\begin{eqnarray}
\hat{\rho}(0)&=&  \left( \begin{array}{cc}
|z\rangle \langle z| & 0  \\
0          & 0  \\
\end{array}   \right)
\end{eqnarray}

We divide the Hamiltonian (2.7) into
a sum of two terms which commute with each other, i.e.,
\begin{equation}
\hat{H} =\hat{H}_o +\hat{H}_I,\hspace{0.3cm} [\hat{H}_o ,\hat{H}_I ]=0
\end{equation}
with
\begin{eqnarray}
\hat{H}_o &=& \omega \left( \begin{array}{cc}
\hat{n} +\frac{1}{2} & 0  \\
0          & \hat{n}-\frac{1}{2}  \\
\end{array}   \right) ,   \hspace{0.1cm}
\hat{H}_I = \left( \begin{array}{cc}
0                       &\lambda \hat{a}  \\
\lambda \hat{a}^+       & 0    \\
\end{array}   \right)
\end{eqnarray}
where $\hat{n}=\hat{a}^{\dagger}\hat{a}$.

Similarly, we can make the following decomposition:
\begin{equation}
\hat{H}^2 =\hat{A} +\hat{B},\hspace{1.5cm} [\hat{A} ,\hat{B} ]=0
\end{equation}
where the representations of the operators $\hat{A}$ and $\hat{B}$
in the two-dimensional atomic basis
take the following forms:
\begin{equation}
\hat{A} =  \left( \begin{array}{cc}
\hat{S}_{n+1}    & 0  \\
0                & \hat{S}_n \\
\end{array}   \right)
\hspace{1.5cm}
\hat{B} =2\lambda \omega  \left( \begin{array}{cc}
0   & \hat{a} (\hat{n}-\frac{1}{2}) \\
(\hat{n}-\frac{1}{2})\hat{a}^+   & 0 \\
\end{array}   \right)
\end{equation}
where
\begin{equation}
\hat{S}_n=\omega^2(\hat{n}-\frac{1}{2})^2 +\lambda^2\hat{n}
\end{equation}
For convenience, we introduce the following auxiliary operator:
\begin{equation}
\hat{\rho}_2 (t) =\exp(\hat{S} t)\exp(\hat{T} t) \hat{\rho} (0)
\end{equation}

{}From the definition of the superoperators and the  initial condition
(2.13), we find that
\begin{equation}
\hat{\rho}_2 (t) =\exp(-i \hat{H}_I t) \exp(-\gamma t \hat{B})
\hat{\rho}_1 (t) \exp(-\gamma t \hat{B})\exp(i\hat{H}_I t)
\end{equation}

Here
the operator $\hat{\rho}_1 (t)$  is defined by
\begin{eqnarray}
\hat{\rho}(0)&=&  \left( \begin{array}{cc}
\mid \Psi (t)\rangle \langle \Psi (t)\mid & 0 \\
0          & 0  \\
\end{array}   \right)
\end{eqnarray}
where
\begin{equation}
\mid \Psi (t)\rangle =\exp\{-\gamma t [\omega^2 (\hat{n} +\frac{1}{2})^2
+ \lambda^2 \hat{a} \hat{a}^+] \} \mid ze^{-i\omega t}\rangle
\end{equation}

For the exponential operators on the rhs of Eq.(2.20) we can find that
\begin{equation}
\exp(-\gamma t \hat{B})= \left(  \begin{array}{cc}
\hat{X}_{n+1}    &-\frac{\hat{Y}_n(t)}{\sqrt{\hat{n}}}\\
-\frac{\hat{Y}_{n+1}(t)}{\sqrt{\hat{n}+1}}  & \hat{X}_n(t)
\end{array}   \right)
\end{equation}
\begin{equation}
\exp(-i \hat{H}_I t)= \left(  \begin{array}{cc}
\hat{C}_{n+1}    &-i \frac{\hat{S}_n(t)}{\sqrt{\hat{n}}}\\
-i\frac{\hat{S}_{n+1}(t)}{\sqrt{\hat{n}+1}}  & \hat{C}_n(t)
\end{array}   \right)
\end{equation}
where
\begin{equation}
\hat{C}_n (t)=\cos (\lambda t \sqrt{\hat{n}}), \hspace{1.3cm} \nonumber
\hat{S}_n (t)=\sin (\lambda t \sqrt{\hat{n}})
\end{equation}

\begin{equation}
\hat{X}_n (t)=\cosh [2\lambda \gamma \omega t (\hat{n}- \frac{1}{2})
\sqrt{\hat{n}}], \hspace{1.3cm} \nonumber
\hat{Y}_n (t)=\sinh [2\lambda \gamma \omega t (\hat{n}- \frac{1}{2})
\sqrt{\hat{n}}]
\end{equation}
Then,
from Eqs.(2.23) and (2.24) it follows that
\begin{eqnarray}
\exp(-i \hat{H}_I t)\exp(-\gamma t \hat{B} )= \left(  \begin{array}{cc}
\hat{R}_{n+1}(t)   &  -\hat{a}\frac{\hat{V}_n(t)}{\sqrt{\hat{n}}} \\
-\hat{a}^+ \frac{\hat{V}_{n+1}(t)}{\sqrt{\hat{n}+1}} & \hat{R}_n(t)
\end{array}   \right)
\end{eqnarray}
where
\begin{equation}
\hat{R}_{n}=\hat{C}_n(t) \hat{X}_n(t) + i \hat{S}_n(t) \hat{Y}_n(t)
\end{equation}
\begin{equation}
\hat{V}_{n}=\hat{C}_n(t) \hat{Y}_n(t) + i \hat{S}_n(t) \hat{X}_n(t)
\end{equation}

 Substituting Eq.(2.27) into Eq.(2.20),
we can obtain an
explicit expression  for the operator $\hat{\rho}_2 (t)$ as follows:
\begin{eqnarray}
\hat{\rho}_2 (t) =\left( \begin{array}{cc}
\hat{\Psi}_{11} (t)  &  \hat{\Psi}_{12} (t)  \\
\hat{\Psi}_{21} (t)  &  \hat{\Psi}_{22} (t)
\end{array} \right)
\end{eqnarray}
where we have used the  following symbol:
\begin{equation}
\hat{\Psi}_{ij} (t)= \mid \Psi_i (t)\rangle \langle  \Psi_j (t) \mid ,
\hspace{0.5cm} (i,j=1,2)
\end{equation}
with
\begin{equation}
\mid \Psi_1 (t) \rangle =\hat{R}_{n+1}(t)|\Psi (t)\rangle ,\hspace{1.0cm}
\mid \Psi_2 (t)=- \frac{\hat{V}_n(t)}{\sqrt{\hat{n}}} \hat{a}^+| \Psi
(t)\rangle
\end{equation}
where $\mid \Psi (t)\rangle $ is given by equation (2.22).

Taking into account the definition of the superoperator $\hat{R}$,
through the action of the operator $\exp(\hat{R}t)$ on the
 operator $\hat{\rho}_2 (t)$ one can obtain the following formal
 solution:
\begin{equation}
\hat{\rho} (t)=\sum^{\infty}_{k=0} \frac{(2\gamma t)^k}{k!}
\hat{H}^k \hat{\rho}_2 (t) \hat{H}^k
\end{equation}
where the operator $\hat{H}$ and $\hat{\rho}_2 (t)$ are given by Eqs.(2.7)
and (2.30), respectively.

Indeed, Eq.(2.33) is  the exact solution of the master equation (2.6) for the
resonant Jaynes-Cummings Hamiltonian (2.7) in the operator form.
Although the form of the solution (2.33) is
pleasant, it is unconvenient in use.
In most cases of practical interest, one needs to know the matrix
elements of the density operator $\hat{\rho} (t)$ in the two-dimensional atomic
basis to calculate expectation values of observables.

In order to find the explicit form of the solution, we need that
\begin{equation}
\hat{H}^k =\left( \begin{array}{cc}
\hat{f}_{n+1}(k)   & \hat{a}\frac{1}{\sqrt{\hat{n}}} \hat{g}_n(k) \\
\hat{a}^+\frac{1}{\sqrt{\hat{n}+1}} \hat{g}_{n+1}(k)  & \hat{f}_n(k)
\end{array}  \right)
\end{equation}
where
\begin{equation}
\hat{f}_n(k)=\frac{1}{2}(\hat{\varphi}^k_n + \hat{\phi}^k_n), \hspace{1.5cm}
\nonumber
\hat{g}_n(k)=\frac{1}{2}(\hat{\varphi}^k_n - \hat{\phi}^k_n)
\end{equation}
with
\begin{equation}
\hat{\varphi}_n=\omega (\hat{n} -\frac{1}{2}) +\lambda \sqrt{\hat{n}},
\hspace{1.0cm}
\hat{\phi}_n=\omega (\hat{n} -\frac{1}{2}) -\lambda \sqrt{\hat{n}}
\end{equation}

For convenience, we define the following matrix:
\begin{equation}
\hat{\cal{M}}^{(k)} (t)=\hat{H}^k\hat{\rho}_2(t)\hat{H}^k
\end{equation}

{}From Eqs.(30) and (2.34) we can obtain its matrix elements as follows:
\begin{eqnarray}
\hat{\cal{M}}^{(k)}_{11} (t)&=&\hat{f}_{n+1}(k)\hat{\Psi}_{11} (t)
\hat{f}_{n+1}(k)
+\hat{a}\hat{g}_n'(k) \hat{\Psi}_{21} (t) \hat{f}_{n+1}(k)    \nonumber  \\
& &+\hat{f}_{n+1}(k)\hat{\Psi}_{12} (t) \hat{g}_n'(k)\hat{a}^+
+ \hat{a}\hat{g}_n'(k) \hat{\Psi}_{22} (t) \hat{g}_n'(k)\hat{a}^+  \\
\hat{\cal{M}}^{(k)}_{22} (t) &=& \hat{g}_n'(k)\hat{a}^+\hat{\Psi}_{11}
(t)\hat{a} \hat{g}_n'(k)
+ \hat{f}_n(k) \hat{\Psi}_{21} (t)\hat{a} \hat{g}_n'(k)    \nonumber  \\
& &+\hat{g}_n'(k)\hat{a}^+ \hat{\Psi}_{12} (t) \hat{f}_n(k)
+ \hat{f}_n(k) \hat{\Psi}_{22} (t) \hat{f}_n(k)   \\
\hat{\cal{M}}^{(k)}_{21} (t)&=& (\hat{\cal{M}}^{(k)}_{12} (t))^+   \nonumber
\\
&=&\hat{g}_n'(k)\hat{a}^+\hat{\Psi}_{11} (t)\hat{a} \hat{f}_{n+1}(k)
+ \hat{f}_n(k) \hat{\Psi}_{21} (t) \hat{f}_{n+1}(k)    \nonumber  \\
& &+\hat{g}_n'(k)\hat{a}^+ \hat{\Psi}_{12} (t) \hat{g}_n'(k)\hat{a}^+
+ \hat{f}_n(k) \hat{\Psi}_{22} (t) \hat{g}_n'(k)\hat{a}^+
\end{eqnarray}
where
\begin{equation}
\hat{g}_n'(k)=\frac{1}{\sqrt{\hat{n}}} \hat{g}_n(k)
\end{equation}

Therefore, the analytic solution of the master equation (2.6) for the Jaynes-
Cummings Hamiltonian (2.7) can be expressed explicitly as follows:
\begin{equation}
\hat{\rho} (t)= \left( \begin{array}{cc}
\sum^{\infty}_{k=0} \frac{(2\gamma t)^k}{k!} \hat{\cal{M}}^{(k)}_{11} (t) &
\sum^{\infty}_{k=0} \frac{(2\gamma t)^k}{k!} \hat{\cal{M}}^{(k)}_{12} (t) \\
\sum^{\infty}_{k=0} \frac{(2\gamma t)^k}{k!} \hat{\cal{M}}^{(k)}_{21} (t) &
\sum^{\infty}_{k=0} \frac{(2\gamma t)^k}{k!} \hat{\cal{M}}^{(k)}_{22} (t)
\end{array} \right)
\end{equation}
Making use of this solution, one can evaluate mean values of operators of
interest.
In the next two sections  we will use it to investigate  various dynamical
properties of the dissipation JCM.


\setcounter{equation}{0}
\section{Dynamical Properties of the Atom}

In this section, we shall study dynamical properties of the atom and discuss
the
influence of the dissipation on them. We will derive analytic expressions of
 the atomic inversion and the dipole momentum of the atom, and calculate the
atomic  entropy.

\subsection{Collapse and revivals of the atomic inversion}

The atomic inversion is defined as the probability of the atom being the
excited state minus the probability of being the ground state, that is
\begin{equation}
W(t) =Tr[\hat{\rho}_A (t)\hat{\sigma}_3]
\end{equation}
where $\hat{\rho}_A(t)$ is the reduced density operator of the atom, it can be
obtained
through tracing over the field part in (2.42). Making use of the solution
(2.42),
one can rewrite the inversion (3.1) as
\begin{equation}
W(t) =\sum^{\infty}_{k,n=0} \frac{(2 \gamma t)^k}{k!}
[\langle n|\hat{\cal M}^{(k)}_{11} |n \rangle - \langle n|\hat{\cal
M}^{(k)}_{22} |n \rangle ]
\end{equation}
We now calculate the two expectation values on the rhs of the above equation.
Making use of Eqs.(2.38), (2.39) and (2.42) we find that
\begin{eqnarray} \hspace{-0.8cm}
\langle n\mid \hat{\cal{M}}^{(k)}_{11} (t)\mid n\rangle &=&(f_{n+1}(k))^2
\mid\psi_1 (n,t)\mid^2  +(g_{n+1}(k))^2  \mid \psi_2 (n+1,t) \mid^2   \nonumber
\\
& & +2Re \{ f_{n+1}(k) g_{n+1}(k) \psi^*_1 (n,t) \psi_2 (n+1,t)\}  \\
\langle n\mid \hat{\cal{M}}^{(k)}_{22} (t) \mid n\rangle &=&(g_n(k))^2
\mid\psi_1 (n-1,t)\mid^2  +(f_n(k))^2 \mid \psi_2 (n,t) \mid^2  \nonumber  \\
& &+ 2Re \{f_n(k) g_n(k) \psi_2 (n,t) \psi^*_1 (n-1,t)\}
\end{eqnarray}
where functions $f_n(k)$ and $g_n(k)$ are given through replacing the number
operator $\hat{n}$ by the number $n$ in Eq.(2.35), and we have introduced the
following symbols:
\begin{equation}
\psi_1 (n,t)= \langle n\mid \Psi_1 (t)\rangle ,  \hspace{1.3cm}
\psi_2 (n,t)= \langle n\mid \Psi_2 (t)\rangle
\end{equation}
which can be explicitly written as
\begin{eqnarray}
\psi_1(n,t)&=&\frac{1}{2} Q_n \left \{C_{n+1}[1+ \exp(4\gamma \lambda \omega
t(n+\frac{1}{2})
\sqrt{n+1})] \right.  \nonumber \\
& &\left.-iS_{n+1} [1- \exp(4\gamma \lambda \omega t(n+\frac{1}{2})
\sqrt{n+1})] \right \}
\cdot\exp(-\gamma t \varphi^2_{n+1}) e^{-in\omega t}   \\
\psi_2(n,t)&=&\frac{1}{2} Q_n \left \{-C_n[1- \exp(4\gamma \lambda \omega
t(n-\frac{1}{2})
\sqrt{n})] \right. \nonumber  \\
& &\left.  +iS_n [1+ \exp(4\gamma \lambda \omega t(n-\frac{1}{2})
\sqrt{n})] \right \}
\cdot\exp(-\gamma t \varphi^2_{n}) e^{-i(n-1)\omega t}
\end{eqnarray}
where  $C_n(t)$, $S_n(t)$, and $\varphi_n$ are obtained from
 their corresponding operator form
through the replacement: \begin{math}\hat{n} \rightarrow n\end{math}.

{}From Eqs.(3.3), (3.4), (3.6), and (3.7) it follows that
\begin{eqnarray}
\langle n|\hat{\cal M}^{(k)}_{11} (t)|n\rangle &=&\frac{1}{4} |Q_n|^2
\left \{\varphi^{2k}_{n+1}\exp[-2\gamma t \varphi^2_{n+1}] +
\phi^{2k}_{n+1} \exp[-2\gamma t  \phi^2_{n+1}] \right. \nonumber \\
& & +2\varphi^k_{n+1}\phi^k_n \cos[2\lambda
t\sqrt{n+1}] \exp[-2\gamma t \varphi_{n+1} \phi_{n+1}] \nonumber \\
& &\left. \cdot\exp[-4\lambda^2 \gamma t(n+1)]
\right \} \\
\langle n|\hat{\cal M}^{(k)}_{22} (t)|n\rangle &=&\frac{1}{4} |Q_{n-1}|^2
\left \{\varphi^{2k}_{n}\exp[-2\gamma t \varphi^2_{n}] +
\phi^{2k}_{n} \exp[-2\gamma t  \phi^2_{n}] \right. \nonumber \\
& & -2\varphi^k_{n+1}\phi^k_n \cos[2\lambda
t\sqrt{n}] \exp[-2\gamma t \varphi_{n} \phi_{n}] \nonumber \\
& &\left. \cdot\exp[-4\lambda^2 \gamma tn]
\right \}
\end{eqnarray}

Substituting Eqs.(3.8) and (3.9) into Eq.(3.2), after summing over $k$,
we arrive at the result:
\begin{equation}
W(t)= \sum^{\infty}_{n=0} \mid Q_n \mid^2 \exp[-4\lambda^2 \gamma t (n+1)]
\cos(2\lambda t \sqrt{n+1})
\end{equation}
which indicates that the revivals of the atomic inversion are destroyed  in the
time evolution due to the appearance of the decay factor $\exp[-4\gamma
\lambda^2 t(n+1)]$
which comes from the contribution of the damping term in the master equation
(2.6). Obviously, when $\gamma$=$0$, the usual result can be recovered.

\subsection{The dipole moment of the atom}

 As is known, the dynamical properties of the atom involve not only the
 atomic inversion but also the knowledge of the dynamics of coherence between
 the two atomic levels, which can be described by the dipole moment defined by
\begin{equation}
D(t)=Tr[\hat{\rho}_A(t)\hat{\sigma}_-]
\end{equation}
which can be expressed explicitly as follows:
\begin{equation}
D(t) =\sum^{\infty}_{k,n=0} \frac{(2 \gamma t)^k}{k!}
\langle n|\hat{\cal M}^{(k)}_{12} |n \rangle
\end{equation}

It is not trival to evaluate the mean value on the rhs of the above equation.
{}From Eq.(2.40) it follows that
\begin{eqnarray}
\langle n\mid \hat{\cal{M}}^{(k)}_{12} (t)\mid n\rangle &=&
    f_{n+1}(k) f_{n}(k) \psi^*_2 (n,t) \psi_1 (n,t) \nonumber \\
& &+f_{n+1}(k) g_{n}(k) \psi^*_1 (n-1,t) \psi_1 (n,t) \nonumber \\
& &+g_{n}(k) g_{n+1}(k) \psi^*_1 (n-1,t) \psi_2 (n+1,t) \nonumber \\
& &+f_{n}(k) g_{n+1}(k) \psi^*_2 (n,t) \psi_2 (n+1,t)
\end{eqnarray}
Taking into account Eq.(2.35), one can express (3.13)  as a more useful form:
\begin{eqnarray}
\langle n|\hat{\cal M}^{(k)}_{12}(t)|n\rangle &=&
\frac{1}{4}\varphi^k_{n+1}\varphi^k_n[\psi^*_1 (n-1,t) \psi_1 (n,t)
                                     +\psi^*_2 (n,t) \psi_1 (n,t) \nonumber \\
& &+\psi^*_1 (n-1,t) \psi_2 (n+1,t)
                                     +\psi^*_2 (n,t) \psi_2 (n+1,t)]
                                     \nonumber\\
& &+\frac{1}{4}\varphi^k_{n+1}\phi^k_n[-\psi^*_1 (n-1,t) \psi_1 (n,t)
                                     +\psi^*_2 (n,t) \psi_1 (n,t) \nonumber \\
& &-\psi^*_1 (n-1,t) \psi_2 (n+1,t)
                                     +\psi^*_2 (n,t) \psi_2 (n+1,t)]
                                     \nonumber\\
& &+\frac{1}{4}\phi^k_{n+1}\varphi^k_n[\psi^*_1 (n-1,t) \psi_1 (n,t)
                                     +\psi^*_2 (n,t) \psi_1 (n,t) \nonumber \\
& &-\psi^*_1 (n-1,t) \psi_2 (n+1,t)
                                     -\psi^*_2 (n,t) \psi_2 (n+1,t)]
                                     \nonumber\\
& &+\frac{1}{4}\phi^k_{n+1}\phi^k_n[-\psi^*_1 (n-1,t) \psi_1 (n,t)
                                     +\psi^*_2 (n,t) \psi_1 (n,t) \nonumber \\
& &+\psi^*_1 (n-1,t) \psi_2 (n+1,t)
                                     -\psi^*_2 (n,t) \psi_2 (n+1,t)]
\end{eqnarray}

Substituting Eq.(3.14) into (3.12) and making use of Eqs.(3.6) and (3.7),
after summing over $k$ and through a lengthy calculation we obtain the result:
\begin{eqnarray}
D(t)&=& \frac{1}{4} \sum^{\infty}_{n=0} Q_n
Q^*_{n-1} \exp\left \{-\gamma t[\omega -\lambda a_-(n)]^2 \right \} \nonumber
\\
& &\cdot \left \{ \exp[4\gamma \lambda \omega t(na_+(n)- \frac{1}{2}a_-(n))]
                  \cdot e^{-i[\lambda a_-(n)+\omega ]t}  \right. \nonumber  \\
              & &\left. -\exp[-4\gamma \lambda \omega t((n+\frac{1}{2})a_-(n)+
\lambda \sqrt{n(n+1)}]
                  \cdot e^{i[\lambda a_+(n)-\omega ]t}  \right. \nonumber  \\
              & &\left. +\exp[4\gamma \lambda \omega t((n-\frac{2}{2})a_-(n)-
\lambda \sqrt{n(n+1)}]
                  \cdot e^{-i[\lambda a_+(n)+\omega]t} \right. \nonumber  \\
              & &\left.-\exp[-4\gamma \lambda \omega t(na_+(n)+
\frac{1}{2}a_-(n))]
                  \cdot e^{-i[\lambda a_-(n)-\omega]t} \right \}
\end{eqnarray}
where
\begin{equation}
a_{\pm}(n)=\sqrt{n} \pm \sqrt{n+1}
\end{equation}

\subsection{The entropy of the atom}
 It is well known that if we assume that initially the atom and the field in
the
 JCM are in a pure state,then at \begin{math}t>0\end{math}, the atom-field
 system evolves into an entangled state. In this entangled state the  field and
 the atom separately are mixed states. Phoenix and  Knight [25] have shown that
 entropy is the most appropriate quantity of measuring the purity of the
 quantum state in the JCM. The time evolution of the atomic (field) entropy
 reflects the time evolution of the degree of entanglement between the atom
 and the field.The higher the entropy is, the greater the entanglement
 between the atom and the field becomes.

The atomic entropy is defined in terms of the reduced density operator of the
atom in this form:
\begin{equation}
S_A(t)=-Tr[\hat{\rho}_A(t)\ln\hat{\rho}_A(t)]
\end{equation}
where the reduced density operator of the atom can be written as
\begin{eqnarray}
\hat{\rho}_A(t)&=&  \left( \begin{array}{cc}
\lambda_{11} &\lambda_{12}  \\
\lambda_{21} &\lambda_{22}   \\
\end{array}   \right)
\end{eqnarray}
where
\begin{equation}
\lambda_{ij} =\sum^{\infty}_{k,n=0} \frac{(2 \gamma t)^k}{k!}
\langle n|\hat{\cal M}^{(k)}_{ij} |n \rangle,  \hspace{2cm} (i,j=1,2)
\end{equation}
It is easy to see that the off-diagonal element $\lambda_{12}$ equals
the diploe moment of the atom, i.e., \begin{math} \lambda_{12}=D(t)\end{math}.
Making use of Eqs.(3.8) and (3.9), from Eq.(3.19) one can find that
\begin{equation}
\lambda_{11}=\frac{1}{2} \sum^{\infty}_{n=0} \mid Q_n \mid^2 \{1+
\exp[-4\lambda^2 \gamma t (n+1)] \cos(2\lambda t \sqrt{n+1}) \}
\end{equation}
\begin{equation}
\lambda_{22}=\frac{1}{2} \sum^{\infty}_{n=0} \mid Q_{n-1} \mid^2 \{1-
\exp[-4\lambda^2 \gamma t n] \cos(2\lambda t \sqrt{n}) \}
\end{equation}

Since the trace is invariant under a similarity transformation, one can go to
a basis  in which the reduced density operator of the atom is diagonal. Then,
the atomic entropy (3.17) can be expressed as follows:
\begin{equation}
S_A(t)=-\alpha_+\ln{\alpha_+} -\alpha_-\ln{\alpha_-}
\end{equation}
where
\begin{equation}
\alpha_{\pm}=\frac{1}{2} \{1 \pm \sqrt{1-4(\lambda_{11}\lambda_{22}
-|\lambda_{12}|^2)} \}
\end{equation}

\setcounter{equation}{0}
\section{Dynamical Properties of the Field}

As is known, the field in the JCM can exhibit a lot of nonclassical effects in
the time evolution, such as oscillations of the photon-number distribution,
quadrature squeezing and sub-Poissonian photon statistics. In this section, we
shall investigate dynamical properties of the cavity field in the presence of
dissipation, and discuss the influence of the dssipation on these nonclassical
effects.

\subsection{Oscillations of the photon-number distribution}

It is easy to show that the probability $p(n,t)$ of finding $n$ photons in the
cavity field is given by
\begin{equation}
p(n,t) =\sum^{\infty}_{k=0} \frac{(2 \gamma t)^k}{k!}
[\langle n|\hat{\cal M}^{(k)}_{11} |n \rangle + \langle n|\hat{\cal
M}^{(k)}_{22} |n \rangle ]
\end{equation}
where two expectation values on the rhs of the above equation have been given
explicitly in the previous section.

Substituting Eqs.(3.8) and (3.9) into (4.1), after summing over $k$ we obtain
\begin{eqnarray}
p(n,t)&=&\frac{1}{2} \mid Q_n \mid^2 \{1+ \exp[-4\lambda^2 \gamma t
(n+1)] \cos(2\lambda t \sqrt{n+1}) \} \nonumber \\
& & +\frac{1}{2} \mid Q_{n-1} \mid^2 \{1- \exp[-4\lambda^2 \gamma t n]
\cos(2\lambda t \sqrt{n}) \}
\end{eqnarray}

With the help of this probability distribution, it is straightforward to
obtain the mean number of photons in the cavity field with the result:
\begin{equation}
\langle \hat{n}(t)\rangle=\bar{n}
+\frac{1}{2}-\frac{1}{2}e^{-\bar{n}}\sum^{\infty}_{n=0}
\frac{\bar{n}^n}{n!} \exp[-4\lambda^2 \gamma t
(n+1)] \cos(2\lambda t \sqrt{n+1})
\end{equation}
where $\bar{n}=|z|^2$ is the initial mean photon number in the
field. As expected, from expressions (4.2) and (4.3) we can see that there is
two decay factors $exp[-4\lambda^2 \gamma t(n+1)]$ and $exp[-4\lambda^2\gamma
tn]$ which come from the damping term in the master equation (2.6). The
oscillatory behaviors of $p(n,t)$ and $\langle \hat{n}(t)\rangle $ are
weakened with increasing of the parameter $\gamma$. Since the damping  is
proportional to the temperature of the environment, these oscillatory
behaviors deteriorate with the increasing of the environmental temperature.
When the damping vanishes, (4.2) and (4.3) reduce to
\begin{equation}
p(n,t)=|Q_n |^2 \cos^2(\lambda t\sqrt{n+1})
+|Q_{n-1}|^2 \sin^2(\lambda t\sqrt{n})
\end{equation}
\begin{equation}
\langle \hat{n}(t)\rangle=\bar{n}
+\frac{1}{2}-\frac{m}{2}e^{-\bar{n}}\sum^{\infty}_{n=0}
\frac{\bar{n}^n}{n!} \cos(2\lambda t\sqrt{n+1})
\end{equation}
which are just the usual expressions without dissipation.

\subsection{Squeezing properties of the cavity field}

We now study the quadrature squeezing of the field in the
cavity field. We introduce the two
slowly varying Hermitian quadrature components of the field $\hat{X}_1$
and $\hat{X}_2$ defined by, respectively,
\begin{equation}
\hat{X}_1=\frac{1}{2}(\hat{a}e^{i\omega t} +\hat{a}^+ e^{-i\omega
t}),\hspace{1.0cm} \hat{X}_2=\frac{1}{2i} (\hat{a}e^{i\omega t}
-\hat{a}^+ e^{-i\omega t})
\end{equation}
The commutation of $\hat{X}_1$ and $\hat{X}_2$ is
$[\hat{X}_1, \hat{X}_2]=\frac{i}{2}$. The variances $\langle(\Delta
X_i)^2\rangle \equiv \langle\hat{X}_i^2\rangle-(\langle\hat{X}_i\rangle)^2$
($i=1,2$) satisfy
the Heisenberg uncertainty relation $\langle(\Delta \hat{X}_1)^2\rangle
\langle(\Delta
\hat{X}_1)^2\rangle \geq \frac{1}{16}$. A state of the field is said to
be squeezed when one of the quadrature components $\hat{X}_1$ and
$\hat{X}_2$ satisfies the uncertainty relation  $\langle(\Delta
\hat{X}_i)^2\rangle < \frac{1}{4}$. The degree of squeezing can be
measured by the squeezing parameters [26]  $S_{i}$
($i=1,2$) defined by
\begin{equation}
S_i=\frac{\langle(\Delta \hat{X}_i)^2\rangle -\frac{1}{2}
|\langle[\hat{X}_1,\hat{X}_2]\rangle|}{\frac{1}{2}|\langle[\hat{X}_1,\hat{X}_2]\rangle| }
\end{equation}
which can be expressed in terms of the annihilation and creation operators of
the field as follows:
\begin{eqnarray}
S_1&=& 2\langle\hat{a}^+ \hat{a}\rangle +2{\bf Re}\langle\hat{a}^2 e^{(i2\omega
t)}\rangle
-4({\bf Re} \langle\hat{a}e^{i\omega t}\rangle)^2\\
S_2&=& 2\langle\hat{a}^+ \hat{a}\rangle -2{\bf Re}\langle\hat{a}^2 e^{(i2\omega
t)}\rangle
-4({\bf Im} \langle\hat{a}e^{i\omega t}\rangle)^2
\end{eqnarray}
 Then, the condition for squeezing in the quadrature component can
simply be written as $S_i<0$.

In principle, expectation values for any function $F(\hat{a}^+, \hat{a})$ are
calculated by the following formula:
\begin{equation} \hspace{-0.5cm}
\langle F(\hat{a}^+, \hat{a}) \rangle=
 \sum^{\infty}_{n=0} \sum^{\infty}_{k=0} (\frac{2\gamma}t)^k{k!}
[\langle n|\hat{\cal M}^{(k)}_{11} (t) F(\hat{a}^+,
\hat{a})|n\rangle
+\langle n|\hat{\cal M}^{(k)}_{22} (t) F(\hat{a}^+,
\hat{a})|n\rangle]
\end{equation}
However, it is generally not an easy matter to calculate expectation value of
an
arbitrary function $F(\hat{a}^+, \hat{a})$ for the field in the
JCM. For the expectation value $\langle\hat{a} e^{i\omega t}\rangle$, through
a tedious calcutation we find that
\begin{eqnarray}
\langle\hat{a}e^{i\omega t}\rangle &=&\frac{1}{4} \sum^{\infty}_{n=0} Q_n
Q^*_{n-1} \exp[\gamma t(S_{n+1}+S_n)] \nonumber  \\
& &\cdot\left \{ a_+(n)T_1 \exp(2\gamma t\varphi_{n+1}\varphi_n)
+a_-(n)T_2 \exp(2\gamma t\varphi_{n+1}\phi_n) \right.  \nonumber \\
&  &\left.+a_+(n)T_3 \exp(2\gamma t\varphi_{n}\phi_{n+1})
+a_+(n)T_4 \exp(2\gamma t\phi_{n+1}\phi_n) \right \}
\end{eqnarray}
where $S_n$, $\varphi_n$, and $\phi_n$  are given by the replacement:
\begin{math}\hat{n} \rightarrow n\end{math} in Eqs.(2.18) and (2.36), and
\begin{eqnarray}
T_1&=&[C_{n+1}(t)-iS_{n+1}(t)][C_n(t)+iS_n(t)]   \nonumber \\
& &\cdot [X_{n+1}(t)-iY_{n+1}(t)][X_n(t)-iY_n(t)]
\end{eqnarray}
\begin{eqnarray}
T_2&=&[C_{n+1}(t)-iS_{n+1}(t)][C_n(t)-iS_n(t)]   \nonumber \\
& &\cdot [X_{n+1}(t)-iY_{n+1}(t)][X_n(t)+iY_n(t)]
\end{eqnarray}
\begin{eqnarray}
T_3&=&[C_{n+1}(t)+iS_{n+1}(t)][C_n(t)+iS_n(t)]   \nonumber \\
& &\cdot [X_{n+1}(t)+iY_{n+1}(t)][X_n(t)-iY_n(t)]
\end{eqnarray}
\begin{eqnarray}
T_4&=&[C_{n+1}(t)-iS_{n+1}(t)][C_n(t)-iS_n(t)] \nonumber \\
& &\cdot [X_{n+1}(t)+iY_{n+1}(t)][X_n(t)+iY_n(t)]
\end{eqnarray}
where the functions $C_n(t)$, $S_n(t)$, $X_n(t)$, and $Y_n(t)$ are given by
the replacement:\begin{math}\hat{n} \rightarrow n \end{math} in Eqs.(2.25) and
(2.26).

Substituting the explicit expressions  of the functions
$C_n(t)$, $S_n(t)$, $X_n(t)$, and $Y_n(t)$ into Eqs.(4.12) $\sim$  (4.15),
from Eq.(4.11) we  find that
\begin{eqnarray}
\langle\hat{a} e^{i\omega t}\rangle &=&\frac{1}{4} \sum^{\infty}_{n=0} Q_n
Q^*_{n-1}\left \{ a_+(n)\exp[i\lambda a_-(n)t]
\exp[-b_-(n)\gamma t] \right.\nonumber \\
& &+a_-(n) \exp[-i\lambda a_- (n)t]\exp[-b_+(n)\gamma t]  \nonumber \\
& &+a_-(n) \exp[i\lambda a_+ (n)t]
\exp[-c_-(n)\gamma t] \nonumber \\
& &\left. +a_+(n) \exp[-i\lambda a_+ (n)t]
\exp[-c_+(n)\gamma t]   \right \}
\end{eqnarray}
where
\begin{equation}
b_{\pm}(n)=[ \omega \pm \lambda a_-(n)]^2, \hspace{1.2cm} \nonumber
c_{\pm}(n)=[ \omega \pm \lambda a_+(n)]^2
\end{equation}

Similarly, we can find that
\begin{eqnarray}
\langle\hat{a}^2 e^{i2\omega t}\rangle &=&\frac{1}{4} \sum^{\infty}_{n=0}
\sqrt{n}Q_nQ^*_{n-2} \exp[-\gamma t(S_{n+1}+S_{n-2})]  \nonumber \\
& &\cdot \{ a'_+(n) T'_1 \exp(2\gamma t
\varphi_{n+1}\varphi_{n-1})
+a'_-(n) T'_2 \exp(2\gamma t\varphi_{n+1}\phi_{n-1})  \nonumber \\
& &+a'_- T'_3 \exp(2\gamma t\varphi_{n-1}\phi_{n+1})
+a'_+(n)T'_4 \exp(2\gamma t\phi_{n+1} \phi_{n-1}) \}
\end{eqnarray}
where
\begin{equation}
a'_{\pm}=\sqrt{n-1} \pm \sqrt{n+1}
\end{equation}
and
\begin{eqnarray}
  T'_1&=&[C_{n+1}(t)-iS_{n+1}(t)][C_{n-1}(t)+iS_{n-1}(t)] \nonumber \\
& &\cdot [X_{n+1}(t)-iY_{n+1}(t)][X_{n-1}(t)-iY_{n-1}(t)]
\end{eqnarray}
\begin{eqnarray}
  T'_2&=&[C_{n+1}(t)-iS_{n+1}(t)][C_{n-1}(t)-iS_{n-1}(t)] \nonumber \\
& &\cdot [X_{n+1}(t)-iY_{n+1}(t)][X_{n-1}(t)+iY_{n-1}(t)]
\end{eqnarray}
\begin{eqnarray}
  T'_3&=&[C_{n+1}(t)+iS_{n+1}(t)][C_{n-1}(t)+iS_{n-1}(t)] \nonumber \\
& &\cdot [X_{n+1}(t)+iY_{n+1}(t)][X_{n-1}(t)-iY_{n-1}(t)]
\end{eqnarray}
\begin{eqnarray}
  T'_4&=&[C_{n+1}(t)+iS_{n+1}(t)][C_{n-1}(t)-iS_{n-1}(t)] \nonumber \\
& &\cdot [X_{n+1}(t)+iY_{n+1}(t)][X_{n-1}(t)+iY_{n-1}(t)]
\end{eqnarray}

Substituting the explicit expressions of the functions
$C_n(t)$, $S_n(t)$, $X_n(t)$, and $Y_n(t)$ into  the above equations,
from Eq.(4.18)  we find that
\begin{eqnarray}
\langle\hat{a}^2 e^{i2\omega t}\rangle
&=&\frac{1}{4} \sum^{\infty}_{n=0} \sqrt{n}Q_n Q^*_{n-2}
\cdot \left \{ a'_+{n} \exp[i\lambda a'_-(n)t]  \exp[-2b'_-(n)\gamma t ]
\right.\nonumber \\
& &+a'_+{n} \exp[-i\lambda a'_-(n)t]\exp[-b'_+(n)\gamma t]  \nonumber \\
& &+a'_-(n) \exp[i\lambda a'_+ (n)t]\exp[-c'_-(n)\gamma t] \nonumber \\
& &\left.+a'_-(n) \exp[-i\lambda a'_+ (n)t]\exp[-c'_+(n)\gamma t]   \right \}
\end{eqnarray}
where
\begin{equation}
b'_{\pm}(n)=[2\omega \pm \lambda a'_-(n)]^2, \hspace{1.0cm} \nonumber
c'_{\pm}(n)=[2\omega \pm \lambda a'_+(n)]^2
\end{equation}

So far we have completed  calculations of all expectation values
needed in the  squeezing parameters Eqs.(4.8) and (4.9). It is
straightforward to obtain the two squeezing parameters $S_1$ and
$S_2$ through the simple substitution of the expectation values
(4.3), (4.11), and (4.24) into Eqs.(4.8) and (4.9).

We now analyze influence  of the dissipation  on the
quadrature squeezing of the cavity field. It is a well-known fact that
without dissipation,the field
in the JCM governed by the von Neumann equation exhibts quardature squeezing
[27,28]. From Eq.(2.6)  we can see that when the damping vanishes,
 it reduce to the coventional von Neumann equation. This means that  in
 the JCM with dissipation there
exists the quadrature squeezing of the field when \begin{math}\gamma
\rightarrow 0\end{math}. Then, taking into account
Eqs.(4.3), (4.11)  and (4.24), from Eqs.(4.8) and (4.9) we can see that
the damping  term in  equation (2.6) leads to the
appearance of decay factors in each term in the  expressions of
the squeezing parameters. Thus,  each term of $S_1$ and $S_2$ decays
with the decrease of the damping parameter $\gamma$. In
particular, at an arbitrary time $t$, \begin{math}\gamma t \gg 1\end{math},
 we find that
\begin{equation}
S_1=S_2 \doteq 2|z|^2+1 > 0
\end{equation}
which means that the quadrature squeezing of the field vanishes
with the increase of the damping. Therefore, we can conclude
that the dissipation suppresses the quadrature
squeezing of the field in the JCM.

\subsection{Sub-Poissonian photon statistics}

A good measure of the extent to which the photon statistics of a state of
the light is sub-Poissonian is the Mandel's $Q$ parameter [29] defined by
\begin{equation}
Q=\frac{\langle\hat{n}^2\rangle-\langle(\Delta \hat{n})^2\rangle}
{\langle\hat{n}\rangle}
\end{equation}
which characterizes the departure from Poissonian photon statistics.
The Mandel parameter  vanishes for a Poissonian distribution. When
\begin{math}Q<0\end{math}, the photons are sub-Poissonian while for
\begin{math}Q>0\end{math} the photons are super-Poissonian.

The expectation value  $\langle\hat{n}\rangle$ in (4.27) has been evaluated in
Eq.(4.3).
And the expectation value $\langle\hat{n}^2\rangle$ can be expressed as
\begin{equation}
\langle\hat{n}^2\rangle=\sum^{\infty}_{n=0} n^2 p(n,t)
\end{equation}

{}From Eqs.(4.2) and(4.28), it follows that
\begin{equation}
\langle\hat{n}^2\rangle=(1+\bar{n})^2
-\frac{1}{2}e^{-\bar{n}} \sum^{\infty}_{n=0} \frac{(2n+1)}{n!}
(\bar{n})^n \exp[-4\lambda^2 \gamma t(n+1)]
\cos[2\lambda t \sqrt{n+1}]
\end{equation}
where $\bar{n}=|z|^2$.

It is straightforward to get the exact expression of
the Mandel's parameter $Q$ through
the substitution of  Eqs.(4.28) and (4.29) into Eq.(4.27).

As is known, when the damping vanishes, i.e., \begin{math} \gamma =0\end{math},
the cavity field in the JCM can exhibit sub-Poissonian distribtuion [30]
in the time evolution governed by the von Neumann equation. Moreover, from
Eqs.(4.3) and (4.29) we can see that for an arbitrary given time $t$, when
\begin{math}\lambda^2\gamma t \gg 1\end{math}, we find that
\begin{equation}
\langle\hat{n}\rangle \doteq \bar{n}+\frac{1}{2} \hspace{2cm}
\langle\hat{n}^2\rangle \doteq (1+\bar{n})^2
\end{equation}
which leads to
\begin{equation}
Q=\frac{4\bar{n} +1}{4\bar{n} +2} > 0
\end{equation}
This indicates that the state  of the cavity field is super-Piossonian.
Therefore, we can conclude that the state of the field in the JCM changes from
sub-Poissonian (or Poissonian) distribution to the super-Poissonian
distribution with the increase of the damping.

\section{Conclusions}
     In this paper we have presented an alternative approach to deal
     with the JCM with the dissipation at finite temperature by
     introducing a new master equation for the reduced density operator
     of the system  under the Born-Markovian approximations. We have
     found the analytic solution of the master eqution for the JCM when
     initially the field is in a coherent state and the atom in its
     excited state. We have studied dynamical properties of the atom
     and the field in the JCM in detail.
     We have obtained analytic expressions of the atomic inversion
     the atomic dipole moment and the atomic entropy.
     We have investigated
influence  of the dissipation on nonclassical effects in the JCM,
such as revivals of   the atomic inversion, oscillations of the photon-number
distribution, the quadrature sqeezing of the cavity field, and sub-Poissonian
photon statistics. In particular, we have shown
 that the dissipation suppresses
     nonclassical effects in the JCM. In our model, the damping $\gamma$
     linearly increases with the environmental temperature, therefore
     these nonclassical effects deteriorate with increasing the
     environmental temperature.

\vspace{0.5cm}
\begin{flushleft} \Large \bf
Acknowledgments
\end{flushleft}
We would like to thank  our colleagues at Theoretical Physics Division,
Nankai Institute of Mathematics for their  useful discussions.
 This research was supported by the National Natural Science
Foundation of China.

\end{document}